\documentclass[showpacs,preprintnumbers,prl,amsmath,amssymb,superscriptaddress, twocolumn]{revtex4-1}
\usepackage{graphicx}
\usepackage{endnotes}
\usepackage{bm}
\usepackage{epsfig}

\newcommand{\etal}{{\it et al.}} 
\newcommand{\TlSe}{TlNi$_{2}$Se$_{2}$}
\newcommand{\TlS}{TlNi$_{2}$S$_{2}$}
\newcommand{\TlSeS}{TlNi$_{2}$SeS}

\begin{document}


\title{Anomalous Pressure Dependence of the Superconducting Transition Temperature in TlNi$_2$Se$_{2-x}$S$_x$}

\author{S. K. Goh}
\affiliation{Department of Physics, The Chinese University of Hong Kong, Shatin, New Territories, Hong Kong, China}
\affiliation{Cavendish Laboratory, University of Cambridge, J. J. Thomson Avenue, Cambridge CB3 0HE, United Kingdom}

\author{H. C. Chang}
\author{P. Reiss}
\author{P. L. Alireza}
\affiliation{Cavendish Laboratory, University of Cambridge, J. J. Thomson Avenue, Cambridge CB3 0HE, United Kingdom}

\author{Y. W. Cheung}

\author{S. Y. Lau}
\affiliation{Department of Physics, The Chinese University of Hong Kong, Shatin, New Territories, Hong Kong, China}

\author{Hangdong Wang}
\affiliation{Department of Physics, Zhejiang University, Hangzhou 310027, China}
\affiliation{Department of Physics, Hangzhou Normal University, Hangzhou 310036, China}
\author{Qianhui Mao}
\affiliation{Department of Physics, Zhejiang University, Hangzhou 310027, China}

\author{Jinhu Yang}
\affiliation{Department of Physics, Hangzhou Normal University, Hangzhou 310036, China}
\author{Minghu Fang}
\affiliation{Department of Physics, Zhejiang University, Hangzhou 310027, China}

\author{F. M. Grosche}
\author{M. L. Sutherland}
\affiliation{Cavendish Laboratory, University of Cambridge, J. J. Thomson Avenue, Cambridge CB3 0HE, United Kingdom}
\date{\today}


\begin{abstract}
We report the pressure dependence of the superconducting transition temperature, $T_c$, in TlNi$_2$Se$_{2-x}$S$_x$ detected via the AC susceptibility method. The pressure-temperature phase diagram constructed for  
TlNi$_{2}$Se$_{2}$, TlNi$_{2}$S$_{2}$ and TlNi$_{2}$SeS exhibits two unexpected features: (a) a sudden collapse of the superconducting state at moderate pressure for all three compositions and (b) a dome-shaped pressure dependence of $T_c$ for TlNi$_{2}$SeS. These results point to the nontrivial role of S substitution and its subtle interplay with applied pressure, as well as novel superconducting properties of the TlNi$_2$Se$_{2-x}$S$_x$ system.

\end{abstract}


\maketitle


The phase diagram of iron-based superconductors often feature a superconducting region in the proximity of antiferromagnetism. The multi-orbital nature of the system and the nesting between well-separated electron and hole Fermi surfaces provide a framework for the discussion of spin-fluctuation-mediated superconductivity as well as spin-density-wave (SDW)  type antiferromagnetism \cite{Mazin08, Kuroki08}.  The 122-type Fe-pnictide family of materials offers a prominent example of these effects \cite{Ishida09, Paglione10, Johnston10, Shibauchi14}, where the application of pressure or chemical substitution can tune the system away from antiferromagnetism and towards superconductivity.  

The situation in the 122-type nickel-based systems is however rather different. The stoichiometric parent compounds already exhibit superconductivity, with BaNi$_2$P$_2$ \cite{Mine08}, BaNi$_2$As$_2$ \cite{Ronning08}, SrNi$_2$P$_2$ \cite{Ronning09} and SrNi$_2$As$_2$ \cite{Bauer08} having comparatively low $T_c$'s of 3~K, 0.7~K, 1.4~K and 0.6~K respectively. The structural transition seen in the Fe-based compounds is of first order in the Ni materials, and may even be absent altogether \cite{Bauer08,Hirai12}. Importantly, magnetic ordering is universally absent in these superconducting materials even with tuning, for example in chemically tuned BaNi$_2$(As$_{1-x}$P$_x$)$_2$ (0 $\leq$ x $\leq$ 0.13) \cite{Kudo12} and pressure tuned BaNi$_2$As$_2$ (up to 27~kbar) \cite{Park10}.  

These observations, taken together with studies of electron-phonon coupling \cite{Subedi08} and Fermi surfaces \cite{Shein09,Ideta14,Zhou11,Terashima09} suggest that the superconductivity in the Ni-based systems is likely of the conventional type, although there remain further compounds to be explored. Understanding the phase diagrams and Fermiology of these materials is an important step in understanding the complex physics of the Fe- and Ni-based pnictides and chalcogenides as a whole, and may help shed light on how the Fe-based pnictides achieve such high $T_c$'s.

\TlSe\ is a relatively new and unstudied member of this family. It superconducts below 3.7~K, and crystallizes in a tetragonal ThCr$_2$Si$_2$-type structure. The normal state is a Pauli paramagnetic metal involving unusually heavy electrons with an effective mass of around 14--20 $m_e$ \cite{Wang13a}. Thermal conductivity measurement suggests that \TlSe\ possesses multiple, nodeless superconducting gaps \cite{Hong14}. The material can be tuned by replacing Se with S, and Wang \etal\ have reported a smooth but non-monotonic variation of $T_c$ as a function of sulphur concentration $x$ in the isostructural series TlNi$_2$Se$_{2-x}$S$_x$ \cite{Wang13b}. Magnetic susceptibility and electrical resistivity measurements in the normal state up to 300~K did not detect any signature of additional phase transition for the entire substitution series, in contract to the closely related system KNi$_2$S$_2$ where a number of structural transitions were observed in the normal state \cite{Neilson13}.

In TlNi$_2$Se$_{2-x}$S$_x$ sulphur substitution is expected to introduce chemical pressure into the system, due to the smaller ionic radius of sulphur. At the same time this process introduces disorder into the system, as quantified by a significant reduction of the residual resistivity ratio (RRR=$\rho_{300{\rm K}}/\rho_{4{\rm K}}$) from $\sim100$ for $x=0$ to less than 10 for $x=1$ and $x=2$. Clearly, the $x$-dependence of $T_c$ not only reflects the effect of chemical pressure, but also the pair-breaking effects of disorder \cite{Wang13b}. The application of hydrostatic pressure on the other hand separates these two effects, and allows us to probe directly the intrinsic pressure dependence of $T_c$.

Single crystals of \TlSe, \TlSeS\ and \TlS\ were synthesized using the self-flux method as described elsewhere \cite{Wang13a, Wang13b}. To track the superconducting transition under pressure, we implemented a two-coil technique in a Moissanite anvil cell, in which a 140-turn modulation coil was wound around the Moissanite anvil and a 10-turn pickup coil was placed inside the gasket hole together with the crystal \cite{Alireza03, Goh08, Klintberg10}. Owing to the advantageous volume filling factor of this setup, typically $\sim$30\%, the signal from the superconducting transition is clear, enabling us to follow the evolution of the superconducting state under pressure accurately. A Helium-3 dipper provided the low temperature environment. To avoid heating, we used a small modulation current of 1~mA, which gave a modulation field of around 1~Oe with a modulation frequency of 1.1~kHz. Glycerin was used as the pressure medium to provide hydrostatic pressure \cite{Osakabe08}, and the pressure was determined via ruby fluorescence spectroscopy.

\begin{figure}[!t]\centering
      \resizebox{8.5cm}{!}{
              \includegraphics{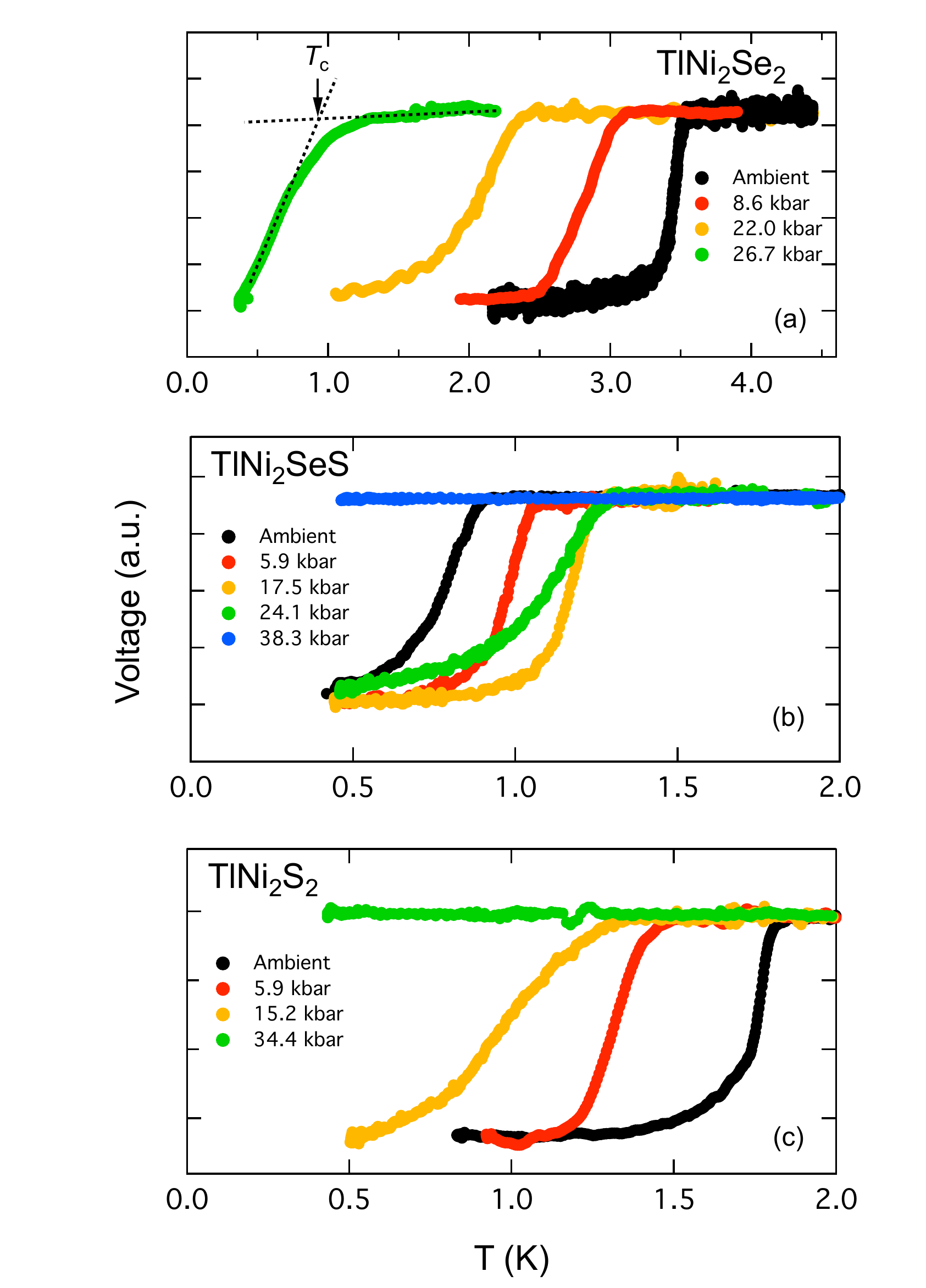}}                				
              \caption{\label{fig1} (Color online) Selected AC susceptibility traces showing the superconducting transitions for (a) \TlSe, (b) \TlSeS\ and (c) \TlS. The dotted lines in (a) illustrates our determination of the superconducting transition temperature.}
\end{figure}

\begin{figure}[!t]\centering
       \resizebox{9cm}{!}{
              \includegraphics{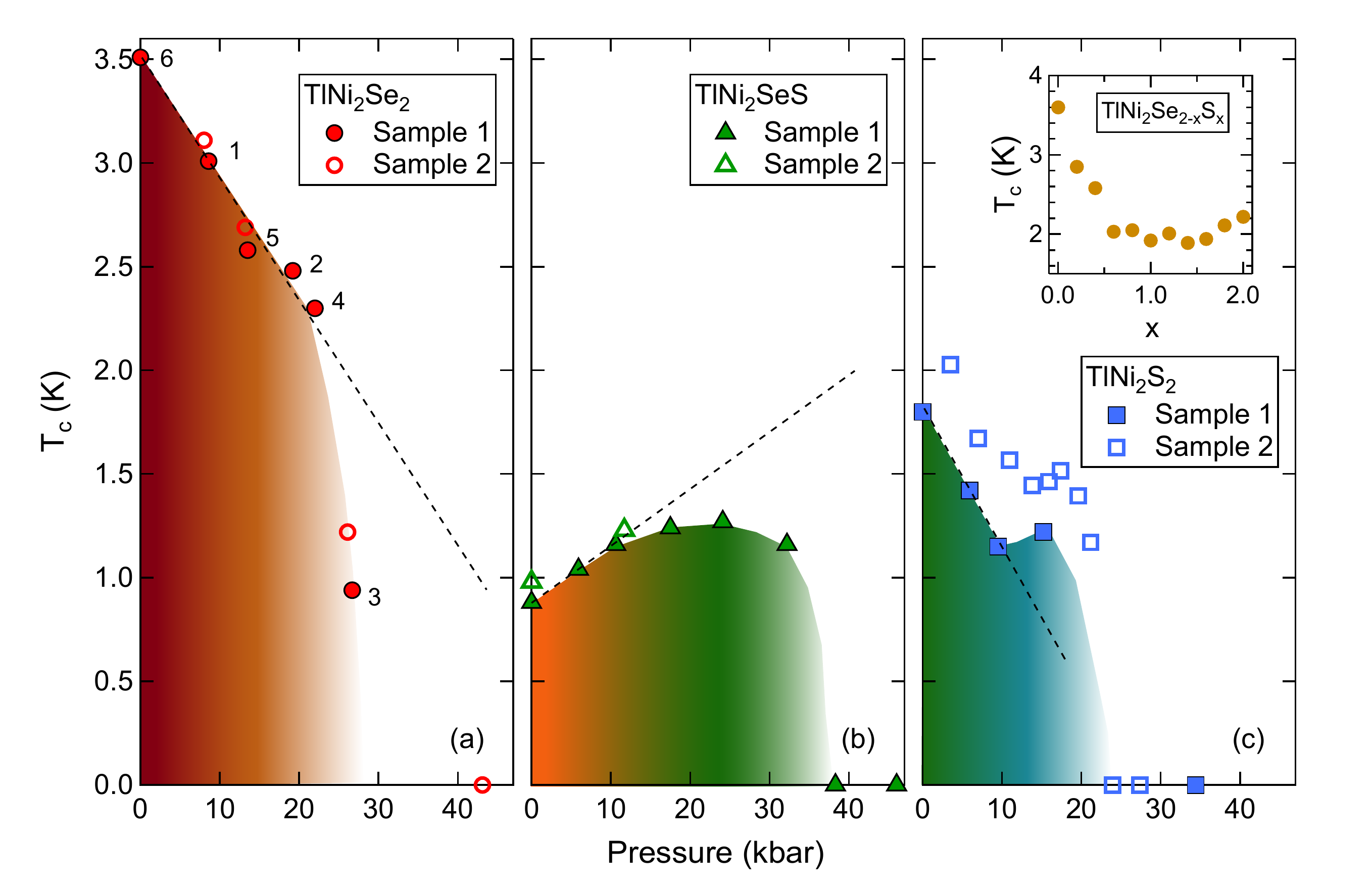}}                				
              \caption{\label{fig2} (Color online) Temperature-pressure phase diagram for (a) \TlSe, (b) \TlSeS\ and (c) \TlS. For each composition, we conducted two sets of measurements to check the reproducibility of the results. Closed and open symbols denoted the first and the second runs, respectively. For \TlSe, the numbers next to the closed circles denote the sequence of the measurements. For the other runs, the measurements were taken in the order of increasing pressures. The inset to (c) shows the $x$ dependence of $T_c$ at ambient pressure in TlNi$_2$Se$_{2-x}$S$_x$ \cite{Wang13b}}
\end{figure}

Fig. \ref{fig1}(a) shows the temperature dependence of the normalised pick-up voltage for \TlSe. This voltage is proportional to the AC susceptibility of the sample, and the transition to the superconducting state is clearly marked by the drop in the pick-up voltage. At ambient pressure, the transition temperature $T_c$, defined as the onset of the transition, is $3.5$~K. This agrees well with the reported value of $3.7$~K determined by resistivity and heat capacity \cite{Wang13a}. With applied pressure, $p$, $T_c$ decreases with an initial slope of d$T_c/$d$p\sim-59$~mK/kbar. This trend continues to about 22~kbar, followed by a rapid suppression of the superconducting state (c.f. Fig. \ref{fig2}a). 

In the isostructural substitution series TlNi$_2$Se$_{2-x}$S$_x$, $T_c$ evolves smoothly as a function of the sulphur content $x$. As shown in the inset to Fig. \ref{fig2}(c), $T_c$ first decreases for $x<1$, and remains more or less constant between $x=1$ and $x=1.6$. For $x>1.6$, $T_c(x)$ shows a gentle positive slope. The replacement of Se by S is accompanied by a monotonic decrease of the lattice parameters $c$ and $a$ \cite{Wang13b}: moving from $x=0$ to $x=2$ results in a $\sim$2\% (5\%) reduction in the lattice parameter $a$ ($c$). If we consider the substitution of S as only providing a chemical pressure to the system, our observation of the collapse of the superconducting state in \TlSe\ at $\sim$22~kbar is unexpected. 

\begin{figure*}[!t]\centering
       \resizebox{17cm}{!}{
              \includegraphics{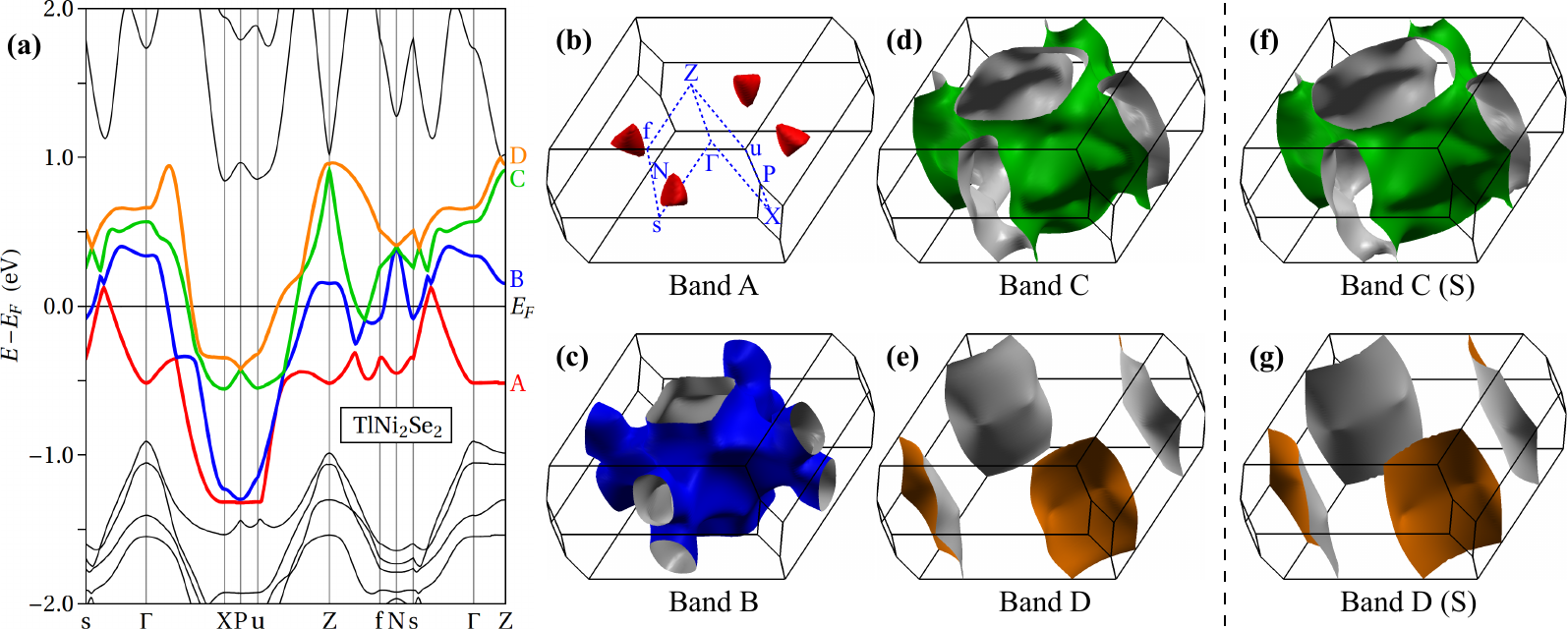}}                				
              \caption{\label{fig3} (Color online) (a) Electronic band structure of \TlSe\ along selected high symmetry directions, giving rise to four Fermi surface sheets as shown in (b) -- (e). Two of the four Fermi surface sheets of \TlS\ are shown in (f) and (g). For the illustrations of the Fermi surface sheets, the $\Gamma$-point is located at the centre, and the $Z$-point is at the top of the Brillouin zone directly above $\Gamma$. The colored side of each Fermi surface corresponds to occupied states.}
\end{figure*}
The pressure dependence of $T_c$ in \TlS\ and \TlSeS\ reveals further unusual features. Figs. \ref{fig1}(b) and \ref{fig1}(c) display the corresponding AC susceptibility data for each of these materials,  allowing us to construct the $T$~--~$p$ phase diagrams as summarized in Fig. \ref{fig2}. If the effect of S substitution were to provide only chemical pressure, the application of physical pressure on fully S-substituted compound, \TlS, would extend the trend of $T_c(x)$ near $x=2$, i.e. $T_c$ would be enhanced under pressure. Instead, $T_c$ decreases rapidly under pressure with a large initial slope of $-68$~mK/kbar. An even more surprising evolution of $T_c(p)$ is observed in \TlSeS: $T_c$ first \textit{increases} with an initial slope  
of $+27$~mK/kbar, reaches a maximum value of $\sim$1.3~K before it plummets, thus forming a dome-shaped $T_c(p)$ frequently observed in many correlated electron systems. 

To check the reproducibility of our results, we conducted two sets of measurements using different samples for each composition. For \TlSe\ and \TlSeS, the agreement between sets of measurements is excellent. For \TlS, the second set (Sample 2) gives an overall higher $T_c(p)$, which we attribute to a slight difference in sample quality. Importantly, the curvature of $T_c(p)$ is identical for both sets of data. In fact, these two sets of data can be brought to a broad agreement by including a relative offset of $\sim$0.4~K. These measurements demonstrate the robustness of our data, and place confidence on our observation of unusual, multi-dome $T_c(p)$ across the TlNi$_2$Se$_{2-x}$S$_x$ series.

A striking similarity across these $T$~--~$p$ phase diagrams concerns the abrupt disappearance of the superconducting state under high pressure. It might be tempting to attribute the disappearance to the onset of pressure inhomogeneity, however we find this unlikely due to previous work using a similar experimental technique. A recent study performed by some of us on BaFe$_2$(As,P)$_2$ up to 70~kbar \cite{Klintberg10} used the same pressure transmitting fluid, same type of anvil cell and similar gasket thickness. However, $T_c(p)$ varied smoothly in BaFe$_2$(As,P)$_2$ up to the highest pressure achieved. Moreover, in the present study the disappearance of superconductivity occurs at different pressures in samples with different sulphur content -- the pressure at which superconductivity collapses is $\sim$25~kbar, 40~kbar and 22~kbar in \TlSe, \TlSeS\ and \TlS, respectively. As a consequence of these facts, it is difficult to associate the disappearance with pressure inhomogeneity.


To gain further insight, we examine the electronic structure of the stoichiometric end compounds \TlSe\ and \TlS. Bandstructure calculations were carried out with the WIEN2k \cite{Wien} package, which is based on density functional theory in the local density approximation. Exchange correlations were approximated with the Perdew-Enzerhoff approximation. Due to the rather heavy cores, additional variational steps accounting for spin-orbit coupling and relativistic local orbits were included. The calculations were performed with a resolution of 100,000 points in the first Brillouin Zone.

We compute the lattice parameters $a$ and $c$ for the fully relaxed structure, and we obtain $a=3.90$~\AA\ and $c=13.56$~\AA\ for \TlSe, in excellent agreement with the experimental values $a=3.87$~\AA\ and $c=13.43$~\AA\ \cite{Wang13b}. For \TlS, the calculated (experimental \cite{Wang13b}) lattice parameters are $a=3.81$~\AA\ ($a=3.79$~\AA) and $c=13.00$~\AA\ ($c=12.77$~\AA). Fig. \ref{fig3}(a) shows the dispersion relation of \TlSe\ along selected high symmetry directions. Four bands cross the Fermi level giving rise to Fermi surface sheets A--D shown in Figs. \ref{fig3}(b)--(e). For \TlS, the electronic structure is very similar to that of \TlSe. In particular, Band A, Band B and Band D assume a very similar Fermi surface topology as the \TlSe\ counterparts. However, it is important to note that the Fermi surface sheet associated with Band C becomes more rounded around the $Z$-point in \TlS. Since the lattice parameters of TlNi$_2$Se$_{2-x}$S$_x$ evolve smoothly with $x$, it is thus very reasonable to assume that the Fermi surfaces of the system would evolve smoothly with $x$. This implies that, for \TlSeS, the Fermi surface sheets associated with Band C would be more (less) rounded at $Z$ than the corresponding sheets in \TlSe\ (\TlS).

The progressive rounding of the Fermi surface sheet associated with Band C will have strong influence on the interband nesting probability with the Fermi surface sheets associated with Band D. The nesting of the well separated hole and electron sheets not only provides an intuitive picture to understand a spin-density-wave type antiferromagnetism, but also present a framework for the discussion of 
spin-fluctuation mediated superconductivity \cite{Mazin08, Kuroki08}. Our calculations thus suggest the tuning of the strength of this nesting condition by varying the sulphur concentration or applying pressures.

The existence of multiple $T_c$ domes has been revealed in several iron-based systems, notably the $T_c(p)$ in FeSe$_{1-x}$ \cite{Miyoshi09, Bendele10} and the $T_c(x)$ in LaFeAsO$_{1-x}$H$_x$ \cite{Iimura12, Hiraishi14}. In these systems, the mechanism of superconductivity was unclear when the multi-dome $T_c$ was first reported \cite{Miyoshi09, Iimura12}. However, subsequent sensitive microscopic measurements detected the existence of antiferromagnetism bordering the superconducting phase \cite{Bendele10, Hiraishi14}, thereby allowing a unified treatment of superconductivity within the framework of spin fluctuation model. Our discovery of the multi-dome $T_c(p)$ in the TlNi$_2$Se$_{2-x}$S$_x$ bears striking resemblance to the phase diagrams of FeSe$_{1-x}$ and LaFeAsO$_{1-x}$H$_x$. With the picture of the Fermi surface nesting from our electronic structure calculations, it is urgently needed to investigate the magnetism of this system under pressure and over a wide range sulphur concentrations.

The clearly different phase diagrams produced through tuning by pressure and chemical substitution in the \TlSe\ system could be accounted for in a number of ways. It may be for instance that the inequivalence arises from the sensitivity of the superconducting state to disorder induced pair-breaking. The most likely pairing scenario for \TlSe\ is a multi-gapped nodeless s$\pm$ wave state, supported by recent thermal conductivity measurements \cite{Hong14}. In this state $T_c$ is expected to be rapidly suppressed with disorder \cite{Onari09}, a result of a breakdown of Anderson's theorem arising from interband scattering between sign-reversing Fermi surface sheets. The disorder introduced through isoelectronic substitution of S for Se could be felt in this manner. A second possibility is that there exists an as-yet-unknown structural or magnetic phase transition that occurs under high pressure. Our results motivate further work on the high pressure structural properties and magnetic properties of this family.

\textbf{Acknowledgment.} We acknowledge funding support from the EPSRC, Trinity College (Cambridge), CUHK (Startup Grant, Direct Grant No. 4053071), UGC Hong Kong (ECS/24300214), Cusanuswerk and the Royal Society. The work in ZJU and HNU was supported by the Natural Science Foundation of China (Grants No. 11374261 and No. 11204059), the Ministry of Science and Technology of China (National Basic Research Program No. 2011CBA00103, No. 2012CB821404, and No. 2015CB921004).


\begin{thebibliography}{34}
\expandafter\ifx\csname natexlab\endcsname\relax\def\natexlab#1{#1}\fi
\expandafter\ifx\csname bibnamefont\endcsname\relax
  \def\bibnamefont#1{#1}\fi
\expandafter\ifx\csname bibfnamefont\endcsname\relax
  \def\bibfnamefont#1{#1}\fi
\expandafter\ifx\csname citenamefont\endcsname\relax
  \def\citenamefont#1{#1}\fi
\expandafter\ifx\csname url\endcsname\relax
  \def\url#1{\texttt{#1}}\fi
\expandafter\ifx\csname urlprefix\endcsname\relax\def\urlprefix{URL }\fi
\providecommand{\bibinfo}[2]{#2}
\providecommand{\eprint}[2][]{\url{#2}}

\bibitem{Mazin08}
I.~I. Mazin, D.~J. Singh, M.~D. Johannes, and M.~H. Du,
\newblock Phys. Rev. Lett. {\bf 101}, 057003 (2008).

\bibitem{Kuroki08}
K.~Kuroki {\em et~al.},
\newblock Phys. Rev. Lett. {\bf 101}, 087004 (2008).

\bibitem{Ishida09}
K.~Ishida, Y.~Nakai, and H.~Hosono,
\newblock J. Phys. Soc. Jpn. {\bf 78}, 062001 (2009).

\bibitem{Paglione10}
J.~Paglione and R.~L. Greene,
\newblock Nature Phys. {\bf 6}, 645 (2010).

\bibitem{Johnston10}
D.~C. Johnston,
\newblock Advances in Physics {\bf 59}, 803 (2010).

\bibitem{Shibauchi14}
T.~Shibauchi, A.~Carrington, and Y.~Matsuda,
\newblock Annu. Rev. Condens. Matter Phys. {\bf 5}, 113 (2014).

\bibitem{Mine08}
T.~Mine {\em et~al.},
\newblock Solid State Commun. {\bf 147}, 111 (2008).

\bibitem{Ronning08}
F.~Ronning {\em et~al.},
\newblock J. Phys.: Condens. Matter {\bf 20}, 342203 (2008).

\bibitem{Ronning09}
F.~Ronning {\em et~al.},
\newblock Phys. Rev. B {\bf 79}, 134507 (2009).

\bibitem{Bauer08}
E.~Bauer, F.~Ronning, B.~Scott, and J.~Thompson,
\newblock Phys. Rev. B {\bf 78}, 172504 (2008).

\bibitem{Hirai12}
D.~Hirai, F.~von Rohr, and R.~Cava,
\newblock Phys. Rev. B {\bf 86}, 100505 (2012).

\bibitem{Kudo12}
K.~Kudo, M.~Takasuga, Y.~Okamoto, Z.~Hiroi, and M.~Nohara,
\newblock Phys. Rev. Lett. {\bf 109}, 097002 (2012).

\bibitem{Park10}
T.~Park, H.~Lee, E.~Bauer, J.~Thompson, and F.~Ronning,
\newblock J. Phys: Conf. Ser. {\bf 200}, 012155 (2010).

\bibitem{Subedi08}
A.~Subedi and D.~Singh,
\newblock Phys. Rev. B {\bf 78}, 132511 (2008).

\bibitem{Shein09}
I.~Shein and A.~Ivanovskii,
\newblock Phys. Rev. B {\bf 79}, 054510 (2009).

\bibitem{Ideta14}
S.~Ideta {\em et~al.},
\newblock Phys. Rev. B {\bf 89}, 195138 (2014).

\bibitem{Zhou11}
B.~Zhou {\em et~al.},
\newblock Phys. Rev. B {\bf 83}, 035110 (2011).

\bibitem{Terashima09}
T.~Terashima {\em et~al.},
\newblock J. Phys. Soc. Jpn. {\bf 78}, 033706 (2009).

\bibitem{Wang13a}
H.~Wang {\em et~al.},
\newblock Phys. Rev. Lett. {\bf 111}, 207001 (2013).

\bibitem{Hong14}
X.~C. Hong {\em et~al.},
\newblock Phys. Rev. B {\bf 90}, 060504 (2014).

\bibitem{Wang13b}
H.~{Wang} {\em et~al.},
\newblock arXiv e-prints  (2013), 1305.1033.

\bibitem{Neilson13}
J.~R. Neilson, T.~M. McQueen, A.~Llobet, J.~Wen, and M.~R. Suchomel,
\newblock Phys. Rev. B {\bf 87}, 045124 (2013).

\bibitem{Alireza03}
P.~L. Alireza and S.~R. Julian,
\newblock Rev. Sci. Instrum. {\bf 74}, 4728 (2003).

\bibitem{Goh08}
S.~K. Goh {\em et~al.},
\newblock Curr. Appl. Phys. {\bf 8}, 304 (2008).

\bibitem{Klintberg10}
L.~E. Klintberg {\em et~al.},
\newblock J. Phys. Soc. Jpn. {\bf 79}, 123706 (2010).

\bibitem{Osakabe08}
T.~Osakabe and K.~Kakurai,
\newblock Jpn. J. of Appl. Phys. {\bf 47}, 6544 (2008).

\bibitem{Wien}
K.~Schwarz and P.~Blaha,
\newblock Computational Materials Science {\bf 28}, 259  (2003).

\bibitem{Miyoshi09}
K.~Miyoshi, Y.~Takaichi, E.~Mutou, K.~Fujiwara, and J.~Takeuchi,
\newblock J. Phys. Soc. Jpn. {\bf 78}, 093703 (2009).

\bibitem{Bendele10}
M.~Bendele {\em et~al.},
\newblock Phys. Rev. Lett. {\bf 104}, 087003 (2010).

\bibitem{Iimura12}
S.~Iimura {\em et~al.},
\newblock Nature Commun. {\bf 3}, 943 (2012).

\bibitem{Hiraishi14}
M.~Hiraishi {\em et~al.},
\newblock Nature Phys. {\bf 10}, 300 (2014).

\bibitem{Onari09}
S.~Onari and H.~Kontani,
\newblock Phys. Rev. Lett. {\bf 103}, 177001 (2009).

  
\end{thebibliography}


\end{document}